\documentclass[11pt]{article}
\usepackage{amsmath,amssymb,color}
 \allowdisplaybreaks
 
 \newcommand{\be}{\begin{equation}}
 \newcommand{\ee}{\end{equation}}
 \newcommand{\bea}{\begin{eqnarray}}
 \newcommand{\eea}{\end{eqnarray}}
 \newcommand{\nn}{\nonumber}
 \newcommand{\td}{\tilde}
 
 \newcommand{\pd}{\partial}
 \newcommand{\cL}{{\cal L}}

 \newcommand{\cO}{{\cal O}}

 \newcommand{\bpsi}{\bar\psi}
\newcommand{\aei}{\it Max-Planck-Institut f\"ur Gravitationsphysik
(Albert-Einstein-Institut)\\
M\"uhlenberg 1, D-14476 Potsdam, Germany}

\newcommand{\auth}{Jianwei Mei}

\begin{document}

%\begin{flushright}
%\hfill{
%MIFP-xxx}\\
%\hfill{
%\bf hep-th/yymmnnn}
%\end{flushright}

\begin{center}

~\vspace{20pt}

\centerline{\Large\bf The Spacetime of a Dirac Fermion}

\vspace{25pt}

%\auth \symbolfootnote[1]{footnote here...}

\auth

%\vspace{10pt}{\ihep}

%\vspace{10pt}{\tamu}

\vspace{10pt}{\aei}

\vspace{25pt}

\underline{ABSTRACT}

\end{center}

We present an approximate solution to the minimally coupled
Einstein-Dirac equations. We interpret the solution as describing
a massive fermion coexisting with its own gravitational field. The
solution is axisymmetric but is time dependent. The metric
approaches that of a flat spacetime at the spatial infinity. We
have calculated a variety of conserved quantities in the system.

%%%%%%%%%%%%%%%%%%%%%%%%%%%%%%
 \newpage
 \setcounter{footnote}{0}
% \setcounter{page}{1}
%%%%%%%%%%%%%%%%%%%%%%%%%%%%%%

\section{Introduction}

Explicit solutions in General Relativity (coupled to matter) and
in supergravity theories play a significant role in the study of
quantum gravity. Several important progresses (for example,
\cite{hawking74,unruh76,strominger.vafa96,maldacena97}) have been
achieved with the help of some concrete solutions. Due to the
weakness of the gravitational interaction, most of the
phenomenologically interesting solutions describe physics at the
astronomical scale, such as black holes, p-branes, black
rings\cite{emparan.reall01} and so on.

From the theoretical point of view, however, it is also
interesting to look for exact solutions at the microscopic scale.
Quantum field theory (QFT) in flat spacetime tells us that if the
zero-point energy gravitates, then extreme fine tuning is needed
to render the cosmological constant to its currently observed
value. There have been a lot of effort towards solving the
problem, but without too much success \cite{weinberg00}. The
zero-point energy is a byproduct of the quantization of fields in
flat spacetime. When it comes to curved spacetime, the situation
is much more complicated (see, e.g. \cite{wald95}). As implied by
the Unruh effect \cite{unruh76}, even the notion of {\it particle}
is no longer fundamental, but depends on the observer. It is our
hope that an exact solution to gravity coupled to quantum matter
may help us better understand the nature of QFT in curved
spacetime, and offer some clue to the persisting problem of the
cosmological constant. When a spacetime is flat at the spatial
infinity, the usual notion of {\it particle} is still useful for
an observer far away from the distribution of matter. One can
imagine a scenario where the observer sees nothing but a single
particle at the center. It is then interesting to ask how the wave
function of the particle (such as a neutrino or electron) behaves
under its own gravity. To find this, we will also need an exact
solution to classical gravity coupled with quantum matter. More
recently, the development in non-Fermi liquid and holographic
superconductors
\cite{liu.mcgreevy.vegh09,hartnoll09,cubrovic.zaanen.schalm09} has
also made it interesting to study gravitational solutions with
back reaction from spinor fields \cite{cubrovic.zaanen.schalm10}.
In this work, we will present an approximate solution to the
Einstein-Dirac system. It describes how a fermion exists, under
its own gravity, in a flat spacetime background.

Compared with the long list of literature on other systems, there
have been relatively fewer work on solutions to the Einstein-Dirac
system (for examples, see \cite{bw57}-\cite{zhelnorovich00}). Most
of the effort came after Brill and Wheeler's 1957 paper
\cite{bw57}. An early review of works focused on neutrinos can be
found in \cite{kuchowicz74}. Some of the solutions describe ghost
spinors (see, e.g. \cite{dr74}). Solutions in dimensions other
than four can be found in \cite{friedrich00,hhz04}. Particle-like
solutions have been previously studied in \cite{fsy98a}.

Apart from the complexity of the calculation, the reason
discouraging people from dealing with the spinor field might be
its anti-commutative nature upon quantization. Here we will treat
the spinor field as a quantum mechanical wave function, but not as
a quantized field operator. The reason is the following. For a
bosonic field, the wave-function treatment can be justified in the
low energy limit, when there is no particle production. On the
other hand, a spinor field is often believed to be intrinsically
quantum, and one should always quantize it first before using it.
However, the problem with gravity is that there is NO known ways
to quantize the spinor field that is completely satisfactory.
There is always a negative energy associated with the un-quantized
spinor field. In the usual QFT treatment, this negativeness is
shifted to hide in the zero-point energy, rendering it physically
irrelevant \cite{peskin.schroeder95}. But in theories with
gravity, the problem reappears in the form of a too large
contribution to the cosmological constant. In this sense, the
cosmological constant problem is essentially the negative energy
problem. So when there is gravity, the usual way of quantizing the
spinor field is no longer a good fix of all the problems. Our hope
is that, an exact particle-like solution to the Einstein-Dirac
system (but with an un-quantized spinor field) may offer some hint
on how to correctly quantize the spinor field with gravity.

Due to the complexity of the coupled Einstein-Dirac equations, we
start by looking for an approximate solution to the system. We
expand the solution in terms of the radial coordinate, and our
solution is approximately valid at the spatial infinity. At the
present stage, it is very difficult to discuss the stability of
the solution.\footnote{I thank the referee for raising up this
point.} In fact, it is not even clear if our approximate solution
comes from a {\it well behaved} full solution or not. On the other
hand, it is already non-trivial to find an ansatz (such as
(\ref{expansion.guv}) and (\ref{expansion.psi})) which allows for
an approximate solution to exist.\footnote{For example, the
counterpart of (\ref{expansion.guv}) and (\ref{expansion.psi})
allow for no solution when there is an arbitrary cosmological
constant.} So as a first step towards finding an exact
particle-like solution to the Einstein-Dirac system, we will be
content with the approximate solution for now, but will try to
solve all the remaining problems in future works.

I will present the solution in next section. Then I will discuss
some of the (approximately) conserved quantities. A short summary
is at the end.

\section{The Solution}

The action of the Einstein-Dirac system is given by
%%%
\be S=\int d^nx\sqrt{|g|}\Big\{\frac{R-2\Lambda}{16\pi G_{(n)}}
-\frac{i}2\bpsi\gamma^\mu D_\mu\psi +\frac{i}2 (D_\mu
\bpsi)\gamma^\mu\psi-i\mu\bpsi\psi\Big\}\,,\label{action.ED}\ee
%%%
where $G_{(n)}$ is Newton's constant in $n$ spacetime dimensions,
$\Lambda$ is the cosmological constant, and $\mu$ is the mass of
the spinor field. The signature of the metric is mostly positive.
It is often convenient to use the Planck mass $M_p^2=\frac1{8\pi
G_{(n)}}$ in place of Newton's constant. We will let $16\pi
G_{(n)}=1$ from now on. The notations related to the spinor field
are
%%%
\bea D_\mu\psi&=&\Big(\pd_\mu+\frac{i}2w_{ab\mu}\gamma^{ab}
\Big)\psi\,,\quad\bpsi=\psi^\dagger\gamma^0\,,\nn\\
D_\mu\bpsi&=&\pd_\mu\bpsi-\frac{i}2w_{ab\mu}\bpsi\gamma^{ab}
\,,\quad\gamma^{ab}=-\frac{i}{4}[\gamma^a~,~\gamma^b]\nn\\
\{\gamma^a\,,\,\gamma^b\}&=&2\eta^{ab}\,,\quad g_{\mu\nu}
=\eta^{ab}e_{a\mu}e_{b\nu}\,,\quad \gamma^\mu
=\gamma^ae_a^{~\mu}\,,\nn\\
w_{ab\mu}&=&(\Gamma^\rho_{\mu\nu}e_{a\rho}-\pd_\mu
e_{a\nu})e_b^{~\nu}=-(\nabla_\mu e_{a\nu})e_b^{~\nu}\,.\eea
%%%
Note $\gamma^0$ in $\bpsi$ is defined in the vielbein basis. The
equations from (\ref{action.ED}) are
%%%
\bea0&=&\gamma^\mu D_\mu\psi+\mu\psi=-(D_\mu\bpsi) \gamma^\mu
+\mu\bpsi\,,\label{eom.psi}\\
R_{\mu\nu}&=&\frac{i}8 \bpsi(\gamma_\mu D_\nu+\gamma_\nu
D_\mu)\psi -\frac{i}8 (D_\mu\bpsi\gamma_\nu +D_\nu\bpsi
\gamma_\mu)\psi+\frac{2}{n-2}\Big(\Lambda +\frac{i}4\mu
\bpsi\psi\Big)g_{\mu\nu}\,.\label{eom.guv}\eea
%%%
We will work in four spacetime dimensions ($n=4$) from now on and
will take $\Lambda=0$. As a matter of choice, the gamma matrices
in the vielbein basis are taken to be
%%%
\be\gamma^0=i\left(\begin{matrix}&{\bf 1}_2\cr {\bf
1}_2&\end{matrix}\right) \,,\quad\gamma^{1,2,3}=i
\left(\begin{matrix}&\sigma^{3,1,2}\cr-\sigma^{3,1,2}
&\end{matrix}\right)\,,\ee
%%%
where ${\bf 1}_2$ is the two-dimensional unit matrix and
$\sigma^{1,2,3}$ are the usual Pauli matrices. The corresponding
charge conjugating operator is ${\cal C}=\gamma^3$.

We are interested in an axisymmetric spacetime. The ansatz for the
metric is given by
%%%
\be ds^2=-f_t^2(dt+f_z d\phi)^2 +f_x^2dx^2 +f_y^2 d\theta^2
+(f_pd\phi-\td{f}_z dt)^2\,,\label{ansatz.guv.tmp}\ee
%%%
where $|f_t|>|\td{f}_z|$ in general. We expect $x$ to be the
asymptotic radial coordinate, $\theta$ the latitudinal angle, $t$
the time and $\phi$ the azimuthal angle. It is convenient to write
the vierbeins as
%%%
\be e^0=f_t(dt+f_zd\phi)\,,\quad e^1=f_xdx\,,\quad e^2=f_y
d\theta\,,\quad e^3=f_pd\phi-\td{f}_z dt\,.\label{vielbein.tmp}\ee
%%%
One can always set $\td{f}_z=0$ by using the following local
Lorentz transformation
%%%
\be(\Lambda^a_{~b})=\frac1{\sqrt{1-(\td{f}_z/f_t)^2}}
\left(\begin{matrix}1&&&\td{f}_z/f_t\cr&0&&\cr&&0&\cr
\td{f}_z/f_t&&&1\end{matrix}\right)\,.\ee
%%%
So (\ref{ansatz.guv.tmp}) and (\ref{vielbein.tmp}) are equivalent
to the following,
%%%
\bea&&ds^2=-f_t^2(dt+f_z d\phi)^2 +f_x^2dx^2 +f_y^2 d\theta^2 +
f_p^2d\phi^2\,,\label{ansatz.guv}\\
&&e^0=f_t(dt+f_z d\phi)\,,\quad e^1=f_xdx\,,\quad e^2=f_y
d\theta\,,\quad e^3=f_pd\phi\,.\eea
%%%
We take the spinor field to be
%%%
\bea\Psi= \left(\begin{matrix} \psi_{1a}+i\psi_{1b}\cr
\psi_{2a}+i\psi_{2b}\cr \psi_{3a}+i\psi_{3b}\cr \psi_{4a}
+i\psi_{4b}\end{matrix}\right)\,,\label{ansatz.psi}\eea
%%%
where all the functions $\psi_{ia}$ and $\psi_{ib}$ (through out
the paper, the index $i=1,\cdots,4$) are real.

Given (\ref{ansatz.guv}) and (\ref{ansatz.psi}), it is difficult
to solve (\ref{eom.psi}) and (\ref{eom.guv}) directly. In this
work, we will focus on a spacetime that is flat at the spatial
infinity ($x\rightarrow+\infty$). Our strategy is to expand the
functions in terms of the radial coordinates $x$ and obtain an
approximate solution to the equations when $x\rightarrow+\infty$.
We will assume that all the unknown functions depend on $x,\theta$
and $t$ only. Our ansatz for the functions are
%%%
\bea&&f_t=1+\frac{t_1}x+\frac{t_2}{x^2}+\cO(\frac1{x^3})\,,\quad
f_z=\frac{z_1}x +\frac{z_2}{x^2}+\cO(\frac1{x^3})\,,\nn\\
&&f_x=1+\frac{x_1}x+\frac{x_2}{x^2}+\cO(\frac1{x^3})\,,\quad f_y=x
+y_0+\frac{y_1}x+\frac{y_2}{x^2}+\cO(\frac1{x^3})\,,\nn\\
&&f_p=\sin\theta\Big[x +p_0+\frac{p_1}x+\frac{p_2}{x^2}
+\cO(\frac1{x^3})\Big]\,,\label{expansion.guv}\\
&&\psi_{ia}=\frac{a_{i1}}x+\frac{a_{i2}}{x^2}+\cO(\frac1{x^3})\,,\quad
\psi_{ib}=\frac{b_{i1}}x+\frac{b_{i2}}{x^2} +\cO(\frac1{x^3})\,,\;
i=1,\cdots,4\,, \label{expansion.psi}\eea
%%%
where $x_{1,2,\cdots}$ , $~t_{1,2,\cdots}$ , $~z_{1,2,\cdots}$ ,
$~y_{0,1,2,\cdots}$ , $~p_{0,1,2,\cdots}$ , $~a_{i1,i2,\cdots}$
and $~b_{i1,i2,\cdots}$ are polynomials of trigonometric functions
of $x$, $\theta$ and $t$. The expansion in (\ref{expansion.guv})
guarantees that (\ref{ansatz.guv}) approaches the metric of a flat
spacetime as $x\rightarrow+\infty$.

Given (\ref{expansion.guv}) and (\ref{expansion.psi}), the leading
contribution to (\ref{eom.psi}) is of the order $\cO(\frac1x)$.
The equations are given by
%%%
\bea\mu b_{31}-(\pd_x-\pd_t)a_{11}=0&,&\quad \mu
a_{31}+(\pd_x-\pd_t)b_{11}=0\,,\nn\\
\mu b_{41}+(\pd_x+\pd_t)a_{21}=0&,&\quad \mu
a_{41}-(\pd_x+\pd_t)b_{21}=0\,,\nn\\
\mu b_{11}+(\pd_x+\pd_t)a_{31}=0&,&\quad \mu
a_{11}-(\pd_x+\pd_t)b_{31}=0\,,\nn\\
\mu b_{21}-(\pd_x-\pd_t)a_{41}=0&,&\quad \mu
a_{21}+(\pd_x-\pd_t)b_{41}=0\,.\label{eom.psi.1}\eea
%%%
These equations can be completely solved by
%%%
\bea a_{i1}&=&a_{i1a}(\theta)\cos(w t)\cos(k x)+a_{i1b}(\theta)\cos(w t)\sin(k x)\nn\\
&&+a_{i1c}(\theta)\sin(w t)\cos(k x)+a_{i1d}(\theta)\sin(w
t)\sin(k x)\,,\label{solution.ai1}\eea
%%%
where $i=1,\cdots,4$ and $w^2=k^2+\mu^2$. $b_{i1}$'s are solved in
terms of $a_{i1}$'s in an obvious way. The leading contribution to
(\ref{eom.guv}) is of the order $\cO(x)$. The equations are given
by
%%%
\be(\pd_x^2-\pd_t^2)y_0 =(\pd_x^2-\pd_t^2)p_0=0
\,.\label{eom.guv.1}\ee
%%%
They do not depend on the spinor field. The solutions are
%%%
\bea y_0&=&y_{0y}(\theta)+y_{01}(\theta,x-t)+y_{02}(\theta,x+t)\,,\nn\\
p_0&=&p_{0y}(\theta)+p_{01}(\theta,x-t)+p_{02}(\theta,x+t)\,.
\label{solution.yp0}\eea
%%%
It is hard to see how can functions of $wt\pm kx$ and $x\pm t$
coexist with each other in the case $w\neq k$. So we will assume
$y_{01}=y_{02}=p_{01}=p_{02}=0$ when $\mu\neq0$ (In fact, our
solution exists only when $\mu\neq0$). Note the leading order
equations (\ref{eom.psi.1}) and (\ref{eom.guv.1}) are independent
of each other.

To the order $\cO(\frac1x)$ of (\ref{eom.guv}), one has equations
that determine $z_1,t_1,x_1,y_1$ and $p_1$, in terms of $a_{i1}$
and $b_{i1}$. The assumption about the functions $z_1,t_1,x_1,y_1$
and $p_1$ impose extra constraints on the structure of $a_{i1}$
and $b_{i1}$. For example, two of the equations are
%%%
\bea\pd_x\pd_yt_1&=&-\frac1{4\sqrt{1-y^2}}\Big[a_{11}b_{21} \pd_x
\ln\Big(\frac{a_{11}}{b_{21}}\Big) +a_{21}b_{11}\pd_x\ln \Big(
\frac{a_{21}}{b_{11}}\Big)\nn\\
&&\qquad\qquad-a_{31}b_{41}\pd_x\ln \Big(\frac{a_{31}}{b_{41}}
\Big) -a_{41}b_{31}\pd_x\ln \Big(\frac{a_{41}}{b_{31}}\Big)
\Big]\,,\label{eom.t1}\\
\pd_y\pd_tx_1&=&-\frac1{4\sqrt{1-y^2}}\Big[a_{11}b_{21} \pd_t\ln
\Big(\frac{a_{11}}{b_{21}}\Big) +a_{21}b_{11}\pd_t\ln \Big(
\frac{a_{21}}{b_{11}}\Big)\nn\\
&&\qquad\qquad-a_{31}b_{41}\pd_t\ln \Big(\frac{a_{31}}{b_{41}}
\Big) -a_{41}b_{31}\pd_t\ln \Big(\frac{a_{41}}{b_{31}}\Big)
\Big]\,.\label{eom.x1}\eea
%%%
Using (\ref{solution.ai1}), one can find that the right hand side
of (\ref{eom.t1}) does not depend on $x$, while that of
(\ref{eom.x1}) does not depend on $t$. To be consistent with the
assumption that $t_1$ and $x_1$ only depend on $x$ or $t$ in the
form of trigonometric functions, the right hand sides of both
(\ref{eom.t1}) and (\ref{eom.x1}) must vanish. This condition
constrains both $t_1,x_1$ and $a_{i1},b_{i1}$.  As a special case
that satisfies this condition, we let
%%%
\be b_{11}\sim a_{21}\,,\quad b_{21}\sim a_{11}\,,\quad b_{31}\sim
a_{41}\,,\quad b_{41}\sim a_{31}\,,\ee
%%%
up to some constant factors. This will reduce (\ref{solution.ai1})
to a much simpler form.

The equations from higher orders of the expansion are more
complicated and we will only summarize some of the main features
here. There are eight equations at each order $\cO(1/x^n)$ of
(\ref{eom.psi}). These equations are just enough to determine
$a_{in}$ and $b_{in}$ in terms of $a_{i,n-1}$, $b_{i,n-1}$,
$x_{n-1}$, $y_{n-1}$, $p_{n-1}$, $t_{n-1}$, $z_{n-1}$ and lower
order functions. On the other hand, the equations at the order
$\cO(1/x^n)$ of (\ref{eom.guv}) determine $x_{n-1}$, $y_{n-1}$,
$p_{n-1}$, $t_{n-1}$ and $z_{n-1}$ in terms of $a_{i,n-1}$,
$b_{i,n-1}$, $x_{n-2}$, $y_{n-2}$, $p_{n-2}$, $t_{n-2}$, $z_{n-2}$
and lower order functions. What's more, there are always two
equations that involve $z_n,t_n,x_n,y_n,p_n$, $a_{in}$ and
$b_{in}$ at each order $\cO(1/x^n)$ of (\ref{eom.guv}), while the
rest only involve lower order functions. These equations can
always be solved together with those from the order
$\cO(1/x^{n+1})$ of (\ref{eom.guv}). Because of the constraints
from equations like (\ref{eom.t1}) and (\ref{eom.x1}), it is not
guaranteed that one can find a self-consistent solution to all
orders in $\frac1x$. We have done the calculation up to the order
$\cO(1/x^2)$ for both (\ref{eom.psi}) and (\ref{eom.guv}), and
still we do not see any true obstacle (other than the tediousness)
to push the calculation to higher orders. This is an encouraging
sign that (\ref{expansion.guv}) and (\ref{expansion.psi}) might be
the right ansatz to give us a consistent solution.

At the moment, we will be content with the solution approximate up
to the order $\cO(1/x^2)$. Even at this stage, the full result is
already very unwieldy and contains many free parameters and
functions. Presumably these free functions and parameters should
be determined by equations from higher orders of the expansion.
But for the purpose of giving an accessible example, we have (not
so rigourously) chosen the parameters and functions in such a way
that the solution has a simpler structure. One cannot expect to
get the correct approximation to some exaction solution (if it
exists) in this way. But the solution so obtained is still
approximately valid in its own right, in the sense that it solves
(\ref{eom.psi}) and (\ref{eom.guv}) up to the order $\cO(1/x^2)$.

For the metric, we find that
%%%
\bea f_t&=&1-\frac{N^2w\cos\theta}{4k\mu x}+\cO(1/x^3)\,,\quad
f_p=\sin\theta f_y\,,\nn\\
f_x&=&1-\frac{N^2w^3\cos\theta}{4k^3\mu x} +\frac{N^2k\sin(2wt-2
\zeta) \sin\theta}{4w^2\mu x^2}+\cO(1/x^3)\,,\nn\\
f_y&=&x+\frac{N^2w(2k^2+w^2)\cos\theta}{4k^3\mu} -\frac{N^2
k\sin(2wt-2\zeta)\sin\theta}{4w^2\mu x}\nn\\
&&+\frac{N^4(2k^2-w^2)\sin(2wt-2\zeta)\sin2\theta}{16k^2w
\mu^2x^2}+\cO(1/x^3)\,,\nn\\
f_z&=& +\frac{N^2\Big[6w\cos(2kx)\sin^2\theta+k\sin(2kx)
\sin(2wt-2\zeta)\sin(2 \theta)\Big]}{8k^2\mu^2x^2}\nn\\
&&-\frac{N^2w\sin(2kx)\sin^2\theta}{2k\mu^2x}+\cO(1/x^3)\,.
\label{solution.metric}\eea
%%%
For the spinor field, we find
%%%
\bea\psi_{1a}&=&\frac{N\sqrt{\sin\theta}}{x\sqrt{w-k}}
\Big[\cos(wt+kx-\zeta)-\frac{N^2(4k-w)w\cos\theta
\cos(wt+kx-\zeta)}{8k^2\mu x}\nn\\
&&\qquad\qquad\qquad\qquad\qquad+\frac{(2k+w)\cot\theta
\sin(wt-kx-\zeta)}{4k(w+k)x}\Big]+\cO(1/x^3)\,,\nn\\
\psi_{1b}&=&\frac{N\sqrt{\sin\theta}}{x\sqrt{w+k}}
\Big[\cos(wt-kx-\zeta)-\frac{N^2(4k+w)w\cos\theta
\cos(wt-kx-\zeta)}{8k^2\mu x}\nn\\
&&\qquad\qquad\qquad\qquad\qquad+\frac{(2k-w)\cot\theta
\sin(wt+kx-\zeta)}{4k(w-k)x}\Big]+\cO(1/x^3)\,,\nn\\
\psi_{2a}&=&\frac{N\sqrt{\sin\theta}}{x\sqrt{w-k}}
\Big[\cos(wt-kx-\zeta)-\frac{N^2(4k-w)w\cos\theta
\cos(wt-kx-\zeta)}{8k^2\mu x}\nn\\
&&\qquad\qquad\qquad\qquad\qquad+\frac{(2k+w)\cot\theta
\sin(wt+kx-\zeta)}{4k(w+k)x}\Big]+\cO(1/x^3)\,,\nn\\
\psi_{2b}&=&\frac{N\sqrt{\sin\theta}}{x\sqrt{w+k}}
\Big[\cos(wt+kx-\zeta)-\frac{N^2(4k+w)w\cos\theta
\cos(wt+kx-\zeta)}{8k^2\mu x}\nn\\
&&\qquad\qquad\qquad\qquad\qquad+\frac{(2k-w)\cot\theta
\sin(wt-kx-\zeta)}{4k(w-k)x}\Big]+\cO(1/x^3)\,,\nn\\
\psi_{3a}&=&-\frac{N\sqrt{\sin\theta}}{x\sqrt{w-k}}
\Big[\sin(wt-kx-\zeta)-\frac{N^2(4k-w)w\cos\theta
\sin(wt-kx-\zeta)}{8k^2\mu x}\nn\\
&&\qquad\qquad\qquad\qquad\qquad+\frac{(2k+w)\cot\theta
\cos(wt+kx-\zeta)}{4k(w+k)x}\Big]+\cO(1/x^3)\,,\nn\\
\psi_{3b}&=&\frac{N\sqrt{\sin\theta}}{x\sqrt{w+k}}
\Big[\sin(wt+kx-\zeta)-\frac{N^2(4k+w)w\cos\theta
\sin(wt+kx-\zeta)}{8k^2\mu x}\nn\\
&&\qquad\qquad\qquad\qquad\qquad+\frac{(2k-w)\cot\theta
\cos(wt-kx-\zeta)}{4k(w-k)x}\Big]+\cO(1/x^3)\,,\nn\\
\psi_{4a}&=&-\frac{N\sqrt{\sin\theta}}{x\sqrt{w-k}}
\Big[\sin(wt+kx-\zeta)-\frac{N^2(4k-w)w\cos\theta
\sin(wt+kx-\zeta)}{8k^2\mu x}\nn\\
&&\qquad\qquad\qquad\qquad\qquad+\frac{(2k+w)\cot\theta
\cos(wt-kx-\zeta)}{4k(w+k)x}\Big]+\cO(1/x^3)\,,\nn\\
\psi_{4b}&=&\frac{N\sqrt{\sin\theta}}{x\sqrt{w+k}}
\Big[\sin(wt-kx-\zeta)-\frac{N^2(4k+w)w\cos\theta
\sin(wt-kx-\zeta)}{8k^2\mu x}\nn\\
&&\qquad\qquad\qquad\qquad\qquad+\frac{(2k-w)\cot\theta
\cos(wt+kx-\zeta)}{4k(w-k)x}\Big]+\cO(1/x^3)\,,
\label{solution.psi}\eea
%%%
where $w^2=k^2+\mu^2$, $N$ is a dimensionless normalization
constant and $\zeta$ is an arbitrary phase. For the results given
above, it is possible to set $\zeta=0$ by a shift in $t$. The
function $\cot\theta$ appearing in (\ref{solution.psi}) indicates
that the solution is divergent at $\sin\theta=0$. This could be a
big problem. But such divergence could also be an artifact of the
expansion in (\ref{expansion.guv}) and (\ref{expansion.psi}). For
example, something like ($a$ and $b$ are constants) $$\frac{a}{b
+x\sin\theta}$$ is obviously regular at $\sin\theta=0$, but it
diverges at $\sin\theta=0$ if firstly expanded around
$x\rightarrow+\infty$. For (\ref{solution.psi}), it is possible to
get rid of the divergence in a similar fashion.

We have found (\ref{solution.metric}) and (\ref{solution.psi}) by
assuming that $a_{in}$ and $b_{in}$ are polynomials of
\{$\cos(kx)$, $\sin(kx)$\} and \{$\cos(wt)$, $\sin(wt)$\} up to
the $(2n-1)$'th power, and $z_n$, $t_n$, $x_n$, $y_n$ and $p_n$
are polynomials of \{$\cos(kx)$, $\sin(kx)$\} and \{$\cos(wt)$,
$\sin(wt)$\} up to the $2n$'th power. The dependence on $\theta$
appears as coefficient functions of the polynomials. This
assignment is inspired by the structure of (\ref{solution.ai1}),
(\ref{solution.yp0}) and that of the equations from different
orders of (\ref{eom.psi}) and (\ref{eom.guv}). After solving
equations up to the order $\cO(1/x^2)$, we find that many
coefficient functions are still undetermined. What's more, we have
also found a lot integration constants in the process. Although
these functions and constants are arbitrary at the order
$\cO(1/x^2)$, one should expect many further constraints to arise
from higher order equations in the expansion. So presumably these
functions and constants should be determined by pushing the
calculation to higher orders, which is currently a daunting task.
In reaching (\ref{solution.metric}) and (\ref{solution.psi}), we
have thrown away most of the functions that do not depend on the
strength of the spinor field (i.e. $N$) in an obvious way. These
functions may be important when one wants to go to higher orders.
As a result, (\ref{solution.metric}) and (\ref{solution.psi}) may
not be the correct approximation to an exact solution of
(\ref{eom.psi}) and (\ref{eom.guv}). On the other hand,
(\ref{solution.metric}) and (\ref{solution.psi}) is still a valid
approximate solution, good to the order $\cO(1/x^2)$.

\section{Conserved Quantities}

From (\ref{eom.psi}), one can derive a conserved current of the
spinor field,
%%%
\be {\cal J}^\mu=\bpsi\gamma^\mu\psi\,,\quad\Longrightarrow\quad
D_\mu {\cal J}^\mu= 0\,.\ee
%%%
If there is a time-like killing vector $\xi$, then one can write
down the probability density of the spinor field as
%%%
\be\rho={\cal J}\cdot\xi\,.\label{density.psi}\ee
%%%
For the energy density of the spinor field, there are two
possibilities. One is by using the energy-momentum tensor,
%%%
\be\rho_1=\xi^\mu\xi^\nu T_{\mu\nu}\,,\quad T_{\mu\nu}=R_{\mu\nu}
-\frac{R}2g_{\mu\nu} \,.\label{density.energy1}\ee
%%%
The other is by using the energy-momentum four-vector,
%%%
\be\rho_2=P\cdot\xi\,,\quad P_\mu=-\frac12\bpsi D_\mu\psi
+\frac12D_\mu\bpsi\psi\,. \label{density.energy2}\ee
%%%
It can be checked that $\nabla_\mu P^\mu=0$. We will see that
$\rho_1$ and $\rho_2$ are quantitatively vastly different.

From (\ref{solution.metric}), one can find an approximate
time-like Killing vector $\xi=\pd_t-\frac{\pd_t y_1}x\pd_x$ at the
spatial infinity $x\rightarrow+\infty$,
%%%
\be\cL_\xi g_{\mu\nu}=\left(\begin{matrix} \cdot&-\frac{\pd_\theta
\pd_ty_1}x &\cdot &-\frac{\pd_t^2y_1}x\cr -\frac{\pd_\theta\pd_t
y_1}x& \frac{2\pd_ty_2}x &\cdot &\cdot \cr \cdot &\cdot &\frac{2
\sin^2\theta\pd_ty_2}x &\cdot \cr -\frac{\pd_t^2y_1}x &\cdot
&\cdot &\cdot\end{matrix}\right) +\cO(\frac1{x^2})\,,\ee
%%%
where $\mu,\nu=\{x,\theta,\phi,t\}$, and $y_1,y_2$ can be read off
(\ref{expansion.guv}) and (\ref{solution.metric}). Now the
probability density of the spinor field (\ref{density.psi}) can be
calculated as
%%%
\be \rho=\frac{M_p^2}2\cdot \frac{4N^2w \sin\theta}{\mu^2x^2}
+\cO(\frac1{x^3})\,.\ee
%%%
Here we have restored the Planck mass to make the dimension of the
density manifestly correct. If we interpret the solution as
describing a particle with its center of mass located at $x=0$,
and if we suppose that the chance of finding the particle far away
from $x=0$ is $\varrho$, then
%%%
\bea \varrho\approx\oint dV\rho\approx\frac{2\pi^2M_p^2
N^2wL}{\mu^2}\,,\quad \Longrightarrow \quad N^2\approx
\frac{\mu^2\varrho}{2 \pi^2 M_p^2 w L}\,,\label{integral}\eea
%%%
where
%%%
\be\oint dV=\int_0^Ldx\int_0^\pi d\theta\int_0^{2\pi}d\phi ~x^2
\sin\theta\,.\ee
%%%
The integral (\ref{integral}) is divergent over the whole space,
so we have introduced a cutoff $L$ to regularize the divergence.
Now if the full wave function is finite near the center, then we
must have $\varrho=1$. This is because the distribution
probability is then dominated by the divergent integral in
(\ref{integral}). We will keep $\varrho$ explicit to cover the
possibility that the wave function may actually diverge at $x=0$.
In this case, a significant portion of the probability may be
distributed near the center.

Let's now turn to the energy of the spinor field. The energy
densities (\ref{density.energy1}) and (\ref{density.energy2}) are
found to be
%%%
\bea \rho_1&\approx&\frac{M_p^2}2\cdot\frac{N^2w\cos\theta
\Big[2\mu+3N^2k\sin(2wt -2\zeta)\sin\theta\Big]}{2k\mu^2x^3}
+\cO(1/x^4)\,,\nn\\
\rho_2&\approx&\frac{M_p^2}2\cdot\frac{4N^2w\sin\theta}{\mu x^2}
+\cO(1/x^3)\,.\eea
%%%
Here we have again restored the Planck mass to make the dimensions
manifestly correct. It is obvious that the two energy densities
are very different. While the nature of $\rho_1$ is not very
clear, we will take $\rho_2$ to be the true energy density of the
spinor field at places far away from the center. In fact, it is
easy to see that
%%%
\be\rho_2=\rho\mu+\cO(1/x^3)\,.\ee
%%%
The contribution of $\rho_2$ to the total energy is
%%%
\be E=\oint dV\rho_2=\varrho\mu\,.\ee
%%%
In the case $\varrho=1$, $\rho_2$ dominates the contribution to
the spinor energy. We see that the total amount is exactly $\mu$.
This is as expected: the solution describes a particle without an
apparent kinetic energy, and so the total energy is nothing but
the mass of the particle. However, there is a puzzle. In the
solution the wave function fluctuates in both space and time, and
the frequency $w$ is apparently larger than the mass. We must have
a non-vanishing wave number ($k\neq0$) for the solution to exist.
But it is still unclear how $k$ is related to the particle mass
$\mu$. One may need to know the full solution to answer this
question.

On the geometry side, the Ricci scalar of (\ref{solution.metric})
is
%%%
\be R=\frac{2N^2k\sin(2wt-2\zeta)\sin\theta}{\mu x^2}
+\cO(1/x^3)\,.\ee
%%%
It is obvious that the curvature can be both positive and
negative, and oscillates with a frequency of $2w$. Since the
spacetime is flat at the spatial infinity, one can use the Komar
formula to calculate the energy stored in the
geometry\cite{komar58,chen.lu.pope05},
%%%
\bea M&=&M_p^2\int_{x=L}\ast d\xi =-\frac{M_p^2\pi^2kN^2L}{\mu}
\sin(2wt-2\zeta)\nn\\
&&\qquad\qquad\qquad=-\frac{\varrho}{2}\frac{k\mu}{w}
\sin(2wt-2\zeta) \,.\label{komar.mass}\eea
%%%
This energy fluctuates in time as well and it also goes negative
for half of the time. We do not have a good explanation to this
result at the moment. For an answer, one may have to better
understand how gravity is coupled to quantum matter.

Similar to (\ref{komar.mass}), one can try to calculate the
angular velocity of the geometry,
%%%
\be J=-\frac{M_p^2}2\int_{x=L}\ast d(\pd_\phi) =\frac{4M_p^2 \pi w
N^2L}{3\mu^2}\cos(2kL)~\approx~0\,.\label{komar.angular}\ee
%%%
It will be interesting to compare this result with the angular
momentum of the spinor field. By analogy with the result in a flat
spacetime, we look at the quantity
%%%
\be S^\mu=\frac{i}2\epsilon^\mu_{~\nu\rho\sigma}\xi^\nu \bpsi
\gamma^{\rho\sigma}\psi\,,\quad{\rm with}\quad
|\epsilon_{\mu\nu\rho\sigma}|=\sqrt{|g|}\,.\ee
%%%
We find that all the components vanishes as in
(\ref{komar.angular}). If this result is also true for the full
solution, then it means that the particle described by the
solution does not have a fixed axis of spin.

\section{Summary}

We have presented an approximate solution to the Einstein-Dirac
system. It solves the coupled Einstein-Dirac equations up to the
order $\cO(1/x^2)$, with $x$ being the radial coordinate. The
solution can be interpreted as describing a single Dirac fermion
coexisting with its own gravitational field. The metric approaches
that of a flat spacetime at the spatial infinity. If one assumes
that the full wave function is everywhere regular in the whole
space, then the total energy in the spinor field is just the mass
of the particle. The energy in the geometry fluctuates in time,
and it is negative for half of the time. For the solution to
exist, we also need a non-vanishing wave number $k$ in the radial
direction. The value of the wave number is undetermined and we
still know very little about its significance.

A natural generalization of the present work is to include a
cosmological constant in the spacetime background. We will leave
this to future works.

\section*{Acknowledgement}

The author thanks Hong Lu and Chris Pope for early discussions on
related topics. This work was supported by the Alexander von
Humboldt-Foundation.

%\newpage

\end{document}